\documentclass[%
amsmath,amssymb, onecolumn
reprint,%
]{revtex4-1}

\usepackage{color}
\usepackage{graphicx}

\newcommand{\partialx}[1]{\frac{\partial #1}{\partial x}}
\newcommand{\partialy}[1]{\frac{\partial #1}{\partial y}}
\newcommand{\partialz}[1]{\frac{\partial #1}{\partial z}}

\newcommand{\ivec}{\/\mathbf i}
\newcommand{\jvec}{\/\mathbf j}
\newcommand{\kvec}{\/\mathbf k}

\begin{document}

\title{Recovering seldom-used theorems of vector  calculus and their application to problems\\ of electromagnetism.}

\author{Antonio \surname{P\'erez-Garrido}}
\email[E-mail me at: ]{antonio.perez@upct.es}
\affiliation{Dpto.~F\'\i sica Aplicada, Campus Muralla del Mar, Universidad Polit\'ecnica de Cartagena. 30202 Cartagena, Murcia (Spain)
}

\date{\today}
\begin{abstract}
In this paper, we  use  differential forms to prove a number of theorems of integral vector calculus that are rarely found in  textbooks. Two of them, as far as the author knows, have not been published before.  Some possible applications to problems in physics are shared including  a general approach for computing net forces and torques on current-carrying loops that yields insights that are not evident from the standard approach.  
\end{abstract}

\maketitle

\section{Introduction}
Integral vector calculus theorems provide powerful tools for solving problems involving electric and magnetic fields. These theorems, such as Gauss's theorem
\begin{equation}
\oint_{\partial V}\mathbf A\cdot d\mathbf s=\int_V\mathbf \nabla\cdot\mathbf A\, dv,
\label{Gauss}
\end{equation}
where $\mathbf A$ is a vector field and $\partial V$ is the boundary surface  of the volume $V$, and Stokes's theorem
\begin{equation}
\oint_{\partial S}\mathbf  A\cdot d\mathbf r
=\int_S (\mathbf \nabla\times\mathbf A)\cdot d\mathbf s,
\label{StokesVector}
\end{equation}
where $\partial S$ is the bounding curve of the surface $S$, 
allow us, for example, to switch between integral and differential version of Maxwell equations.

Other vector integral theorems are much less known, due, likely, to their lower relevance. Even so, they can be useful in understanding electromagnetic fields. Three examples are the integral identities
\begin{equation}
\oint_{\partial S} \mathbf A\times d\mathbf r=\int_S\left[(\mathbf \nabla\cdot\mathbf A)d\mathbf 
s-\mathbf \nabla(\mathbf A\cdot d\mathbf s)\right]=-\int_S \left(d\mathbf s\times\mathbf \nabla\right)\times\mathbf A,
\label{int_Fxdr}
\end{equation}

\begin{equation}
\oint_{\partial S} fd\mathbf r=-\int_S\mathbf \nabla f \times d\mathbf s,  
\label{int_fdr}
\end{equation}
and
\begin{equation}
\oint_{\partial S}\left(\mathbf C\cdot d\mathbf r\right) \mathbf A=\int_S   d\mathbf s \cdot \left(\mathbf \nabla \times\mathbf C-\mathbf C\times \mathbf \nabla\right)\mathbf A,
\label{int_rdrF}
\end{equation}
for any scalar field $f$ and vector fields  $\mathbf A$ and $\mathbf C$. These theorems are rarely found in textbooks about vector calculus. Furthermore, using \eqref{int_fdr} and \eqref{int_rdrF} and the expression for the vector triple product, as detailed later, we also derive a sort of a corollary:
\begin{equation}
\oint_{\partial S} \mathbf C\times \left(\mathbf A\times d \mathbf r\right) =\int_S\mathbf \nabla \left(\mathbf C\cdot\mathbf A\right)\times d\mathbf s
+\int_S\left[d\mathbf s\cdot\left(\mathbf \nabla \times\mathbf C-\mathbf C\times\mathbf\nabla\right)\right]\mathbf A.  \label{int_rxFxdr}  
\end{equation}

In all these identities,  the orientation of surfaces and curves is crucial. This is a standard issue in vector integral calculus thus in subsequent discussions we will not make further mention about this subject, but it is implicit that one must be cautious when doing calculations.

This work has two purposes.  The first is to show how differential forms can be used to easily demonstrate the identities in Eqs.(\ref{int_Fxdr}-\ref{int_rxFxdr}).  This is done in Sect.~II, which can be skipped by readers who are not already familiar with differential forms.  The second purpose is to show how these identities can be applied to problems in electromagnetism to reveal new insights that can be beneficial for teaching physics to advanced  undergraduate students, which is done in Sect.~III.

\section{Demonstration of identities}
Differential forms are a powerful framework in theoretical physics as demonstrated, for instance, in their application to quantum mechanics in a paper by Hoshino\cite{hoshino1978}. Additionally, differential forms demonstrate their robustness  for vector analysis and for teaching and understanding Maxwell's equations and the principles of electromagnetism~\cite{amar1980,schleifer1983,fumeron2020,warnick2014}.  
In this section, we  employ differential forms as tools for demonstrating identities (\ref{int_Fxdr}-\ref{int_rxFxdr}). 
Readers who are inspired to learn more about differential forms in order to understand these demonstrations are urged to consult several  excellent introductions to the subject~\cite{spivak1965,needham2021,dray2014}.

In this paper, we restrict to $\mathbb R^3$. 
Let $A$ be a 1-form, $dx$, $dy$, $dz$ and  $\ivec$, $\jvec$, $\kvec$  orthonormal bases for 1-forms and vectors, respectively, so a 1-form $A$ can be written as a linear combination of 1-form basis 
\begin{equation}
A=A_xdx+A_y dy+ A_z dz=\mathbf A\cdot d\mathbf r,    
\end{equation}
where 
\begin{equation}
\mathbf A =A_x\,\ivec+ A_y\,\jvec+A_z\,\kvec\end{equation}
and
\begin{equation}
d\mathbf r=dx\,\ivec+dy\,\jvec+dz\,\kvec. 
\label{dr}
\end{equation}
Henceforth, the symbol $\wedge$ is not written between elements of the basis of 1-forms; we will write, for example, $dxdy$ as a short for $dx\wedge dy$, thus \begin{equation}
dxdy=-dydx \,\,\,{\rm and}\,\,\, dxdx=dydy=dzdz=0.
\end{equation}

We are going to use the Stokes's theorem that, formally, says: let $M$ be an oriented $n$-dimensional smooth manifold with boundary $\partial M$ and let's suppose we have an $(n-1)$-form $\alpha$ defined on $M$, the integral of its exterior derivative, $d\alpha$, over $M$ is equal to the integral of $\alpha$ over the boundary, $\partial M$, of $M$:
\begin{equation}
\int_{\partial M}\alpha=\int_M d\alpha.   
\label{Stokes}
\end{equation}

The steps to follow for all demonstrations are very simple: compute the exterior derivative of the integrands of the left hand side of Eqs.~(\ref{int_Fxdr}-\ref{int_rxFxdr})  and apply \eqref{Stokes},
in other words, we are going to show that those integrands  are the potentials for the integrands of the right hand side of their respective equations.

\subsection{Identity \eqref{int_Fxdr}} 
We can write $\mathbf A\times d\mathbf r$ as a  differential form,
\begin{equation}
\mathbf A\times d\mathbf r=\left(A_ydz-A_zdy\right)\ivec  
+\left(A_zdx-A_xdz\right)\jvec+
\left(A_xdy-A_ydx \right)\kvec.
\label{vector_1form}
\end{equation}
Actually, (\ref{vector_1form}) is a vector-valued 1-form. We only have to calculate the exterior derivative of (\ref{vector_1form}) to complete the proof.
\begin{equation}
\begin{split}
d(\mathbf A\times d\mathbf r)=& \left[\left(\partialx{A_y}dx+
\partialy{A_y}dy\right)dz-\left(\partialx{A_z}dx+\partialz{A_z}dz\right)dy\right]\ivec+ \\
&\left[\left(\partialy{A_z}dy+
\partialz{A_z}dz\right)dx-\left(\partialx{A_x}dx+\partialy{A_x}dy\right)dz\right]\jvec+\\
&\left[\left(\partialx{A_x}dx+
\partialz{A_x}dz\right)dy-\left(\partialy{A_y}dy+\partialz{A_y}dz\right)dx\right]\jvec.
\label{derivative}
\end{split}
\end{equation}
We are using the already mentioned fact that after the exterior derivative, the product between differential forms is the anti-symmetric exterior product so 
\begin{equation}
dxdx=dydy=dzdz=0.    
\end{equation}
Regrouping terms in \eqref{derivative} we have
\begin{equation}
\begin{split}
d(\mathbf A\times d\mathbf r)=& \left[ \left(\partialx {A_x}+\partialy{A_y}+\partialz{A_z}\right)dydz-\left(\partialx{A_x}dydz+\partialx{A_y}dzdx+\partialx{A_z}dxdy\right)  \right]\ivec +\\
& \left[ \left(\partialx {A_x}+\partialy{A_y}+\partialz{A_z}\right)dzdx-\left(\partialy{A_x}dydz+\partialy{A_y}dzdx+\partialy{A_z}dxdy\right)  \right]\jvec+\\
& \left[ \left(\partialx {A_x}+\partialy{A_y}+\partialz{A_z}\right)dxdy-\left(\partialz{A_x}dydz+\partialz{A_y}dzdx+\partialz{A_z}dxdy\right)  \right]\kvec,
\end{split}
\end{equation}
which can be written in a more compact form  as
\begin{equation}
d(\mathbf A\times d\mathbf r)=  (\mathbf \nabla\cdot\mathbf A)d\mathbf 
s-\mathbf \nabla(\mathbf A\cdot d\mathbf s),
\label{dFXR}
\end{equation}
where $d\mathbf s$ is a vector valued two-form representing the differential surface element, given by: 
\begin{equation}
d\mathbf s=dydz\,\ivec+dzdx\,\jvec+dxdy\,\kvec. 
\label{ds}
\end{equation}
If we use the vector triple product identity, the relation \eqref{dFXR} can be also written as
\begin{equation}
d(\mathbf A\times d\mathbf r)= -\left(d\mathbf s\times\mathbf \nabla\right)\times\mathbf A.    
\end{equation}
Using the generalized Stokes's theorem \eqref{Stokes} we arrive to find
\begin{equation}
\oint_{\partial S} \mathbf A\times d\mathbf r=\int_S d\left( \mathbf A\times d\mathbf r\right),
\end{equation}
or 
\begin{equation}
\oint_{\partial S} \mathbf A\times d\mathbf r=-\int_S
\left(d\mathbf s\times\mathbf \nabla\right)\times\mathbf A, 
\end{equation}
i.e., equation \eqref{int_Fxdr}. This theorem can be also demonstrated applying Stokes's theorem~\eqref{StokesVector} to vector field $\mathbf C=\mathbf B\times \mathbf A$ for a constant vector $\mathbf B$ and using a number of vector identities~\cite{lass1950}.

\subsection{Identity \eqref{int_fdr}}
This identity is the most widely known of (\ref{int_Fxdr}-\ref{int_rxFxdr}) and can be  found in some textbooks.
Following the same  procedure than before, we express $fd\mathbf r$ as:

\begin{equation}
fd\mathbf r=fdx\ivec+fdy\jvec+fdz\kvec.   
\label{fdr}
\end{equation}

Subsequently, we calculate the exterior derivative of \eqref{fdr}:

\begin{equation}
\begin{split}
d(fd\mathbf r)=&\left(\partialy fdy+\partialz fdz\right)dx\ivec+
\left(\partialx fdx+\partialz fdz\right)dy\jvec+ 
\left(\partialx fdx+\partialy fdy\right)dz\kvec\\
=&\left(\partialz fdzdx-\partialy fdxdy\right)\ivec+
\left(\partialx fdxdy-\partialz fdydz\right)\jvec+
\left(\partialy fdydz-\partialx fdzdx\right)\kvec\\
=&-\mathbf \nabla f\times d\mathbf s,
\end{split}
\end{equation}
where we evaluate the differential area element, again,  as in expression \eqref{ds}. Thus, we have that $d(f d\mathbf{r}) = -\mathbf{\nabla}f \times d\mathbf{s}$. This outcome
leads us to Eq.~\eqref{int_fdr} upon applying the generalized Stokes's theorem \eqref{Stokes}.
\begin{equation}
 \oint_{\partial S} fd\mathbf r=\int_S d\left(fd\mathbf r\right) \Longrightarrow 
 \oint_{\partial S} fd\mathbf r=-\int_S
 \mathbf{\nabla}f \times d\mathbf{s}.
\end{equation}

As already stated~\cite{lass1950,hsu1984}, the theorems (\ref{StokesVector}--\ref{int_fdr}) can be written  in  a unique and  compact expression
\begin{equation}
\oint_{\partial S} d\mathbf r*\mathbf g=
\int_S  \left(d\mathbf s\times \mathbf \nabla\right)* \mathbf g,  
\end{equation}
where the asterisk ($*$) denote dot, cross and ordinary products, respectively. In the latter case $\mathbf g$ is, actually, a scalar $g$. 

\subsection{Identity \eqref{int_rdrF}}
As before, we calculate the exterior derivative of $(\mathbf C\cdot d\mathbf r)\mathbf A$ and apply \eqref{Stokes}. Firstly, we develop the expression as a vector valued differential form:
\begin{equation}
(\mathbf C\cdot d\mathbf r)\mathbf A=\left(C_xdx+C_ydy+C_zdz\right)\left(A_x\mathbf i+A_y\mathbf j+A_z\mathbf k\right).
\end{equation}
For our purposes it is enough to make the exterior derivative of $x$ component and extend the result to the other two components,
\begin{equation}\begin{split}
 d\left[\left(C_xdx+C_ydy+C_zdz\right)A_x\right]=&
 C_x\left(\frac{\partial A_x}{\partial y}dydx+\frac{\partial A_x}{\partial z}dzdx\right)\\
 &+A_x\left(\frac{\partial C_x}{\partial y}dydx+\frac{\partial C_x}{\partial z}dzdx\right)  \\
 &+C_y\left(\frac{\partial A_x}{\partial x}dxdy+\frac{\partial A_x}{\partial z}dzdy\right)\\
 &+A_x\left(\frac{\partial C_y}{\partial x}dxdy+\frac{\partial C_y}{\partial z}dzdy\right)\\
&+C_z\left(\frac{\partial A_x}{\partial x}dxdz+\frac{\partial A_x}{\partial y}dydz\right)\\
&+A_x\left(\frac{\partial C_z}{\partial x}dxdz+\frac{\partial C_z}{\partial y}dydz\right),
\label{eqlong}
\end{split}
\end{equation}
and, after regrouping terms, the right hand side of \eqref{eqlong} reduces to

\begin{equation}
\begin{split}
&\left[\left(C_y\frac{\partial }{\partial x}-C_x\frac{\partial }{\partial y}\right)dxdy 
+\left(C_z\frac{\partial }{\partial y}-C_y\frac{\partial }{\partial z}\right) dydz
+\left(C_x\frac{\partial }{\partial z}-C_z\frac{\partial }{\partial x}\right) dzdx\right.\\
+&\left. \left(\frac{\partial C_y}{\partial x}-\frac{\partial C_x}{\partial y}\right)dxdy 
+\left(\frac{\partial C_z}{\partial y}-\frac{\partial C_y}{\partial z}\right) dydz
+\left(\frac{\partial C_x}{\partial z}-\frac{\partial C_z}{\partial x}\right)dzdx\right]A_x.
\label{drdrF}
\end{split}
\end{equation}
For the other two component we only have to change $A_x$ in \eqref{drdrF} with the corresponding component $A_y$ or $A_z$. Thus, we can check that
\begin{equation}
d\left((\mathbf C\cdot d\mathbf r)\mathbf A\right)=
\left(d\mathbf s \cdot\left(\mathbf\nabla \times \mathbf C-\mathbf C\times \mathbf \nabla\right)\right) \mathbf A,
\end{equation}
where $d\mathbf s$ is given by \eqref{ds} and $d\mathbf s \cdot\left(\mathbf C\times \mathbf \nabla\right)$ is the 2-form scalar operator given by
\begin{equation}
d\mathbf s \cdot\left(\mathbf C\times \mathbf \nabla\right)=\left(C_x\frac{\partial}{\partial y}-C_y\frac{\partial }{\partial x}\right)dxdy
+\left(C_y\frac{\partial }{\partial z}-C_z\frac{\partial }{\partial y}\right) dydz
+\left(C_z\frac{\partial }{\partial x}-C_x\frac{\partial }{\partial z}\right) dzdx.
\end{equation}

Thus, appealing again to Stokes theorem~\eqref{Stokes},
\begin{equation}
\oint_{\partial S} \left(\mathbf C\cdot d\mathbf r\right) \mathbf A=\int_S  d\left((\mathbf C\cdot d\mathbf r)\mathbf A\right),   
\end{equation}
and, after substitution, we arrive
\begin{equation}
\oint_{\partial S} \left(\mathbf C\cdot d\mathbf r\right) \mathbf A =\int_S
\left(d\mathbf s \cdot\left(\mathbf\nabla \times \mathbf C-\mathbf C\times \mathbf \nabla\right)\right) \mathbf A.
\end{equation}

\subsection{Corollary \eqref{int_rxFxdr}}
Using the vector triple product expression we develop the integrand in the left side of \eqref{int_rxFxdr}
 as
\begin{equation}
\mathbf C\times\left(\mathbf A\times d\mathbf r\right)=\left(\mathbf C\cdot d\mathbf r \right)\mathbf A-\left(\mathbf C\cdot\mathbf A\right)d\mathbf r.
\label{corollary_demo}
\end{equation}
For the first term in the right hand side of \eqref{corollary_demo} we employ \eqref{int_rdrF} and, for the second term, \eqref{int_fdr}, thus completing the demonstration of corollary \eqref{int_rxFxdr}. 
The author has been unable to find either \eqref{int_rdrF} or \eqref{int_rxFxdr} in any published document, paper, or textbook.

\section{Application of theorems to problems in electromagnetism}
In this section, we apply the established identities, (\ref{int_Fxdr}--\ref{int_rxFxdr}), to address problems in electromagnetism.  By applying these identities, we aim to unravel some aspects of the physics that may remain obscured when employing conventional approaches.

\subsection{Calculation of the magnetic force on a closed current}

The magnetic force, $d\mathbf{F}$, acting on an infinitesimal segment $d\mathbf{r}$, given by \eqref{dr}, of a loop carrying current $I$, can be expressed as:
\begin{equation}
d\mathbf{F} = Id\mathbf{r}\times\mathbf{B}.
\label{force_infinitesimal_current}
\end{equation}
For a macroscopic current, the net force is determined by summing up all the infinitesimal forces,
\begin{equation}
\mathbf F=I\oint_{\partial S} d\mathbf r\times \mathbf B,
\label{FLoop}
\end{equation}
where $\partial S$ is the closed curve defining the current loop, and the boundary of the surface $S$.
Applying the first equality in Eq.~\eqref{int_Fxdr} yields:
\begin{equation}
\mathbf F =I\int_S \mathbf \nabla(\mathbf B\cdot d\mathbf s), 
\label{F_vector}
\end{equation}
where the term involving  $\mathbf{\nabla \cdot B}$ has been omitted because the magnetic field is divergence free and $d\mathbf s$ is given by \eqref{ds}. We can expand \eqref{F_vector} to 

\begin{equation}
\begin{split}
\mathbf F=I\int_S &\left[\left(\partialx{B_x}dydz+\partialx{B_y}dzdx+\partialx{B_z}dxdy\right)\ivec \right.\\
+&\left(\partialy{B_x}dydz+\partialy{B_y}dzdx+\partialy{B_z}dxdy\right)\jvec\\
+&\left.\left(\partialz{B_x}dydz+\partialz{B_y}dzdx+\partialz{B_z}dxdy\right)\kvec\/\right].
\end{split}
\label{netForce}
\end{equation}
In a uniform magnetic field, all the derivatives of $\mathbf B$ are zero, and therefore the net force, $\mathbf F$, over the loop is zero. We also see that the net force will have a non-zero component along an axis only if a component of the magnetic field is not uniform in that direction. 
 Let see some particular cases of application of \eqref{netForce}. 

\begin{figure}[ht]
\includegraphics[width=8cm]{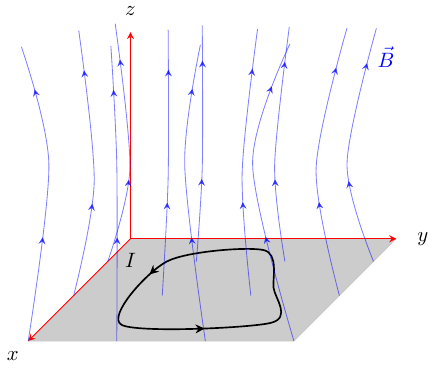}
\caption{\label{planarloop} A planar current loop in a nonuniform magnetic field. We choose a frame of reference with the $z$ axis orthogonal to the loop and  with the $xy$ plane coincident with the loop. }
\end{figure}

 \subsubsection{Planar loop} In this case, we can select a frame of reference with its $z$ axis orthogonal to the plane of the loop, as depicted in Fig.~\ref{planarloop}. The expression \eqref{netForce} reduces to
 \begin{equation}
\mathbf F=I\int_S\left(\partialx{B_z}\mathbf i+\partialy{B_z}\mathbf j+\partialz{B_z}\mathbf k\right)dxdy,
\label{onedirection}
 \end{equation}
as $dydz=dzdx=0$. Consequently, $B_x$ and $B_y$  do not affect  the net force acting on the current loop.
In the  particular case that $B_z$ only depends on $z$,
\begin{equation}
\partialx{B_z}=\partialy{B_z}=0,\,\,\, \partialz{B_z}\ne 0,
\end{equation}
we find the counter intuitive expression
\begin{equation}
\mathbf F=I\int_S\partialz{B_z}\mathbf kdxdy,
\label{counterintuitive}
\end{equation}
indicating the emergence of a net magnetic force parallel to the $z$ axis generated by non-uniformities in $B_z$ along the $z$ axis. This seems to contradict \eqref{FLoop}, where variations along $z$ axis are not taken into account since the integral, in this case, is carried out for a fixed $z$. This apparent contradiction is clarified when we use the fact that the magnetic field is divergence-free, thus 
\begin{equation}
 \partialz{B_z}=-\left(\partialx{B_x}+\partialy{B_y}\right),  
\end{equation}
and \eqref{counterintuitive} is given now by
\begin{equation}
\mathbf F=-I\int_S \left(\partialx{B_x}+\partialy{B_y}\right)  \mathbf kdxdy.   
\end{equation}

Now consider the case of a planar loop placed in the $xy$ plane, as before, but with $B_z$  depending linearly on coordinates:  
\begin{equation}
B_z=b_xx+b_yy+b_zz.  
\end{equation}
After applying \eqref{onedirection} we have a magnetic force given by
\begin{equation}
 \mathbf F=IS\left(b_x\mathbf i+b_y\mathbf j+b_z\mathbf k\right),  
\end{equation}
where $S$ is the area of the surface enclosed by the loop.

\begin{figure}[ht]
\includegraphics{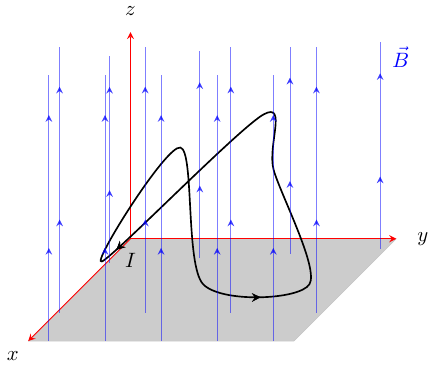}
\caption{\label{fixeddirection} A current loop with an arbitrary shape in a magnetic field along the $z$ direction. The density of field lines may vary depending on the $x$ and $y$ coordinates.}
\end{figure}
\subsubsection{Magnetic field with a fixed direction}
If the magnetic field points only along the $z$-direction, see Fig.~\ref{fixeddirection}, we obtain once again the expression \eqref{onedirection}, which is now valid even for a non planar current loop.
Since $B_x=B_y=0$, the terms $dydz$ and $dzdx$ are discarded, although they can have non-zero values in the general case, as the loop may not be flat. 
In this particular case we have
\begin{equation}
\partialz{B_z}=0,   
\end{equation}
since the magnetic field is divergence free. We are left with the expression
\begin{equation}
\mathbf F=I\int_S\left(\partialx{B_z}\mathbf i+\partialy{B_z}\mathbf j\right)dxdy,
\end{equation}
where we can observe that the magnetic force will be perpendicular 
to the magnetic field and in the direction of magnetic field non-uniformities.

In the particular case that the magnetic field has a linear dependence on coordinates:
\begin{equation}
B_z=b_xx+b_yy,    
\end{equation}
the magnetic force has the form
\begin{equation}
\mathbf F=IS_{\rm net}(b_x\mathbf i+b_y\mathbf j),    
\end{equation}
where $S_{\rm net}$ is the projected area of the surface enclosed by the loop onto the $xy$ plane.
In general, $S_{\rm net}$ is the given by:
\begin{equation}
S_{\rm net}=\int_S\mathbf k\cdot d\mathbf{s}.
\end{equation}
When the enclosed surface is flat,  $S_{\rm net}=\mathbf k\cdot \mathbf S$,
where $\mathbf S$ is the vector representing the surface enclosed by the loop, i.e., a vector orthogonal to the surface and with a magnitude equal to the area of the surface.

\subsection{Torque on a current-carrying loop}
In textbooks, the torque on a flat rectangular electric current loop under the influence of a uniform magnetic field is usually calculated and then generalized so that it can be applied to any loop. This generalization consists of decomposing the current loop into a sum of infinitesimal current loops, so that the rectangular loop formula is valid for any of those  infinitesimal loops.
Here instead we derive the general form of this expression for a uniform magnetic field $\mathbf B$, regardless of the shape of the loop.

We start be integrating the torque about the origin experienced by an infinitesimal segment of the loop.  The force over an infinitesimal current is given by \eqref{force_infinitesimal_current} and
its torque or moment of magnetic force, $\mathrm{d}\mathbf{M}$, about the origin of the frame of reference, is
\begin{equation}
\mathrm{d}\mathbf{M} = \mathbf{r}\times(I\mathrm{d}\mathbf{r}\times\mathbf{B}) = I\mathbf{r}\times(\mathrm{d}\mathbf{r}\times\mathbf{B}),
\end{equation}
where $\mathbf{r}$ is the position vector of the current segment. 
Using the corollary \eqref{int_rxFxdr} in the case of a uniform magnetic field, the total magnetic torque on an arbitrary current loop can be written
\begin{equation}
\begin{split}
\mathbf M&=I\oint_{\partial S} \mathbf r\times\left(d\mathbf r\times \mathbf B\right)=-I\oint_{\partial S} \mathbf r\times\left(\mathbf B\times d\mathbf r\right)   \\
&=-I\int_S\mathbf \nabla \left(\mathbf r\cdot\mathbf B\right)\times d\mathbf s-
I\int_S\left[d\mathbf s\cdot\left(\mathbf \nabla\times\mathbf r-\mathbf r\times\mathbf\nabla\right)\right]\mathbf B.
\label{torque}
\end{split}
\end{equation}
Taking into account that the magnetic field is uniform and that $\mathbf \nabla \times \mathbf r=0$, the torque reduces to
\begin{equation}
\begin{split}
\mathbf M&=-I\int_S\mathbf \nabla \left(\mathbf r\cdot\mathbf B\right)\times d\mathbf s\\
&=-I\int_S \mathbf \nabla (xB_x+yB_y+zB_z)\times d\mathbf s \\
&= I\int_S d\mathbf s\times \left(B_x\ivec+B_y\jvec+B_z\kvec\right)=I\left(\int_{S}d\mathbf s\right)\times \mathbf B.
\end{split}
\end{equation}
That, for a planar loop becomes
\begin{equation}
\mathbf M=I\mathbf S\times\mathbf B,
\label{magneticmoment}
\end{equation}
where $\mathbf S$ is the vector representing the surface enclosed by the loop. Equation \eqref{magneticmoment} is usually written as 
\begin{equation}
\mathbf M=\mathbf{m}\times \mathbf B,   
\end{equation}
where $\mathbf m=I\mathbf S$ receives the name of magnetic moment.

\begin{figure}[ht]
\includegraphics[width=5cm]{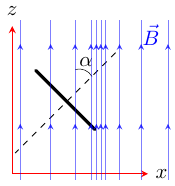}
\caption{\label{planarfixeddirection}
A planar loop in a non uniform magnetic field with a fixed direction. We adopt a reference frame with the $z$-axis aligned parallel to the magnetic field, and the $xz$ plane perpendicular to the planar loop. In this configuration, the loop is visualized as a singular line. The angle $\alpha$ represents the inclination between the magnetic field and the normal to the loop. 
} 
\end{figure}

An interesting case occurs when considering a planar loop exposed to a non uniform magnetic field but with a fixed direction, Fig~\ref{planarfixeddirection}. In this situation, we can select a frame of reference with the $z$ axis aligned with the magnetic field direction, and with the $xz$ plane perpendicular to the loop. The angle between the normal to the loop and the direction of $\mathbf B$ is denoted $\alpha$.
Similar to the previous case, the magnetic field is dependent solely on the variables $x$ and $y$:
\begin{equation}
\mathbf B=B_z(x,y)\mathbf k.    
\end{equation}
Using \eqref{torque} and after performing some algebraic manipulations, we derive the expression for the magnetic torque about the origin of the reference frame as follows:

\begin{equation}
\begin{split}
\mathbf M=I\int_S  &\left\{
z\partialy{B_z}\cos(\alpha)\mathbf i
+\left[B_z\sin(\alpha)-z\partialx{B_z}\cos(\alpha)\right]\mathbf j\right. \\
&\,+\left. \left[
\left(x\partialy{B_z}-y\partialx{B_z}\right)\cos(\alpha)-2z\partialy{B_z}\sin(\alpha)
\right]\mathbf k
\right\} ds,
\end{split}
\label{torqueforplanar}
\end{equation}
where we have $dxdy=\cos(\alpha) ds$ and $dydz=\sin(\alpha) ds$. 
It's worth noting that, in this case, the torque about a  point it is no longer independent of the position of that point.

When the loop is orthogonal to the magnetic field, $\alpha=0$, the torque is given by
\begin{equation}
\mathbf M=I\int_S  \left[
z\partialy{B_z}\mathbf i
-z\partialx{B_z}\mathbf j
+ \left(x\partialy{B_z}-y\partialx{B_z}\right)
\mathbf k
\right] dxdy.
\label{torqueforplanarOrthogonal}
\end{equation}
Eq.~\eqref{torqueforplanarOrthogonal} indicates that $x$ and $y$ components of the torque depend on the nonuniformities of the magnetic field in the $y$ and $x$ directions, respectively, while the $z$ component depends on both.

On the other hand, when the loop is parallel to the magnetic field, $\alpha=90^{\rm o}$, the expression \eqref{torqueforplanar}, becomes
\begin{equation}
\mathbf M=I\int_S  
\left(B_z\,\mathbf j
+2z\partialy{B_z}
\mathbf k\right)
 dydz,
\label{torqueforplanarParallel}
\end{equation}
demonstrating that the torque now lacks a component in the $x$ direction, and the $y$ component remains independent of the location chosen as the point about which the torque is calculated.

Eqs.~\eqref{torqueforplanarOrthogonal} and \eqref{torqueforplanarParallel} provide good qualitative insights into how the shape of the magnetic field can influence the magnetic torque over the current loop, and give hints about how these identities  can offer a comprehensive framework for analyzing the behavior of current loops in different situations.

\section{Conclusions}
We have presented and demonstrated several theorems, (\ref{int_Fxdr}--\ref{int_rxFxdr}), which are not commonly found in standard textbooks on the subject. Notably, a couple of these theorems, namely \eqref{int_rdrF} and \eqref{int_rxFxdr}, have not been found by the author in existing publications. Hence, it is plausible that this paper marks the initial publication of these particular identities. The practical utility of these theorems has been demonstrated through their application to  problems in electromagnetism, where it is shown how the utilization of these theorems can offer a more profound understanding of the physics, shedding light on aspects that may remain hidden when employing more conventional approaches.

\section{Acknowledgements}
I would like to express my gratitude to Antonio Arocas for his meticulous review of the manuscript and his invaluable assistance in researching prior publications of these theorems in the literature on the subject. I also acknowledge support from the grant 
PID2020-120052GB-I00 financed by MCIN/AEI/10.13039/501100011033.


\bibliographystyle{unsrt}

\bibliography{biblio.bib}

\end{document}